\documentclass{article}
\usepackage{spconf}
\usepackage{times}
\usepackage{soul}
\usepackage{url}
\usepackage{xspace}
\usepackage{siunitx}
\usepackage[table]{xcolor}
\usepackage[hidelinks]{hyperref}
\usepackage[utf8]{inputenc}
\usepackage[small]{caption}
\usepackage{graphicx}
\usepackage{amsmath}
\usepackage{amsthm}
\usepackage{booktabs}
\usepackage{multirow}
\usepackage[switch]{lineno}
\usepackage{cleveref}
\usepackage{tabularx}
\usepackage{array}
\usepackage{subcaption}
\usepackage{enumitem}
\usepackage{pifont}
\usepackage{colortbl} \usepackage{bibspacing}
\newcolumntype{C}{>{\centering\arraybackslash}X}
\crefname{section}{§}{§§}
\Crefname{section}{§}{§§}
\ninept

\makeatletter
\renewcommand{\paragraph}[1]{  \par\vspace{0.3em}  \noindent\textbf{#1}\quad}
\makeatother

\usepackage[utf8]{inputenc} \usepackage[T1]{fontenc}    \usepackage{hyperref}       \usepackage{url}            \usepackage{booktabs}       \usepackage{amsfonts}       \usepackage{nicefrac}       \usepackage{microtype}      \usepackage{xcolor}         \usepackage{lipsum}
\usepackage{todonotes}
\usepackage{graphicx}
\usepackage{subcaption}
\usepackage{multirow}
\usepackage{tabularx}
\usepackage{amsmath}
\usepackage{cleveref}
\usepackage{siunitx}
\usepackage{makecell}
\usepackage{array}
\usepackage{float}
\usepackage{tablefootnote}
\usepackage{arydshln}
\usepackage{comment}
\usepackage[ruled,vlined]{algorithm2e}

\newcolumntype{H}{>{\setbox0=\hbox\bgroup}c<{\egroup}@{}}

\usepackage{pifont}
\makeatletter
\newcommand\circled[1]{  \begingroup
        \count@=#1\relax
    \ifnum\count@<1 \count@=1\fi
    \ifnum\count@>10 \count@=10\fi
        \advance\count@ by 171
    \ding{\the\count@}  \endgroup}
\makeatother

\newcommand{\ourmethod}{CodecSlime}

\title{{\Large \textsc{\ourmethod}}: Temporal Redundancy Compression of Neural Speech Codec 
via Dynamic Frame Rate
}

\name{Hankun Wang, Yiwei Guo, Chongtian Shao, Bohan Li, $^{\dagger}$Kai Yu \thanks{$^{\dagger}$\scriptsize{Kai Yu is the corresponding author.}}}
\address{
X-LANCE Lab, School of Computer Science, Shanghai Jiao Tong University, China \\
MoE Key Lab of Artificial Intelligence, Jiangsu Key Lab of Language Computing, China \\
\small{\texttt{\{wanghankun, kai.yu\}@sjtu.edu.cn}}}

\begin{document}
\maketitle
\begin{abstract}
Current mainstream neural speech codecs are fixed-frame-rate (FFR).
However, speech is inherently non-uniform in temporal information density. 
As a result, many tokens are wasted on steady-state segments like long vowels and silences.
To address this mismatch, we present {\ourmethod}, a plugin-style method for compressing temporal redundancy through supporting dynamic frame rate (DFR) on neural speech codecs.
Our method is unsupervised and architecture-agnostic, combining two key innovations, ScheDFR and Melt-and-Cool, for adapting inference and training, respectively.
When integrated into a typical VQ-GAN codec backbone and operating at \SI{40}{Hz} DFR ($\approx$\SI{600}{bps}),  the reconstruction WER of {\ourmethod} is reduced by up to 32\% relative to conventional FFR baselines with the same model architecture and similar bitrates, while other metrics are also competitive. 
{\ourmethod} also enables flexible trade-offs between reconstruction quality and bitrate: 
a single model supports inference at multiple frame rates and consistently outperforms FFR models at the corresponding frame rates.
Audio samples and more results are available at \href{https://x-lance.github.io/codecslime/}{\small\texttt{https://x-lance.github.io/codecslime/}}.
\end{abstract}


\section{Introduction}

The neural speech codec technique, originating from waveform compression~\cite{zeghidour2021soundstream, encodec}, has become widely used in text-to-speech (TTS)~\cite{valle}, speech language models~\cite{borsos2023audiolm} and interaction models~\cite{kyutai2024moshi}, due to its strong compatibility with the discrete tokenization and auto-regressive modeling paradigms of large language models (LLMs)~\cite{brown2020language, guo2025recent}. 
Unlike self-supervised representations~\cite{hsu2021hubert, baevski2020wav2vec}, which typically focus on specific aspects of speech or particular downstream tasks, codecs serve broader application scenarios, for they aim to reconstruct speech signals with the best possible quality at the lowest achievable frame rate and bitrate.

Recent efforts on speech codecs include improving model structure or scaling up model size~\cite{xin2024bigcodec,wu2024ts3codec}, decoupling global information~\cite{ticodec, wang2025spark}, and fusing more semantic information~\cite{zhang2024speechtokenizer,cosyvoice2}, etc.
However, existing codecs still operate at bitrates far exceeding the fundamental theoretical limits (\SI{50}{bps}$\sim$\SI{100}{bps}) derived from phoneme-level statistics~\cite{denes1963statistics} and speech's lexical information bottleneck~\cite{van2017information}.
A fundamental limitation stems from the mismatch between mainstream fixed-frame-rate (FFR) codecs and speech's inherently non-uniform temporal information density~\cite{dieleman2021variable}: 
FFR codecs allocate equal frames to both information-rich and redundant audio segments as long as they have the same time duration. 
SNAC~\cite{Siuzdak_SNAC_Multi-Scale_Neural_2024} and LLM-Codec~\cite{yang2024uniaudio15} adopt hierarchical schemes, assigning different frame rates to distinct quantizer layers, but each layer still operates at a fixed rate. 
Some dynamic frame rate (DFR) approaches have been attempted in previous work on discrete speech tokens, characterized by non-uniform frame durations in the encoded representation, focusing on spoken unit discovery~\cite{cuervo2022variable} or semantics distillation~\cite{baade2024syllablelm, cho2024sylber}. 
However, they fail to produce general-purpose acoustic tokens suitable for reconstruction and rely on overly complex training/inference procedures with limited effectiveness. 
Recently, TFC~\cite{zhang2025unlocking} explored dynamic frame-rate compression for codecs, but it only operates at a fixed target downsampling rate. 
Moreover, its measure of speech dynamics is based on directly computing temporal entropy at the signal level, which limits its ability to capture deeper temporal redundancies. 
Consequently, flexible yet efficient methods for compressing temporal redundancy in neural codecs remain largely unexplored.

\begin{figure}[tb]
  \centering
  \includegraphics[width=0.695\textwidth, trim=0cm 11.6cm 15cm 0cm, clip]  {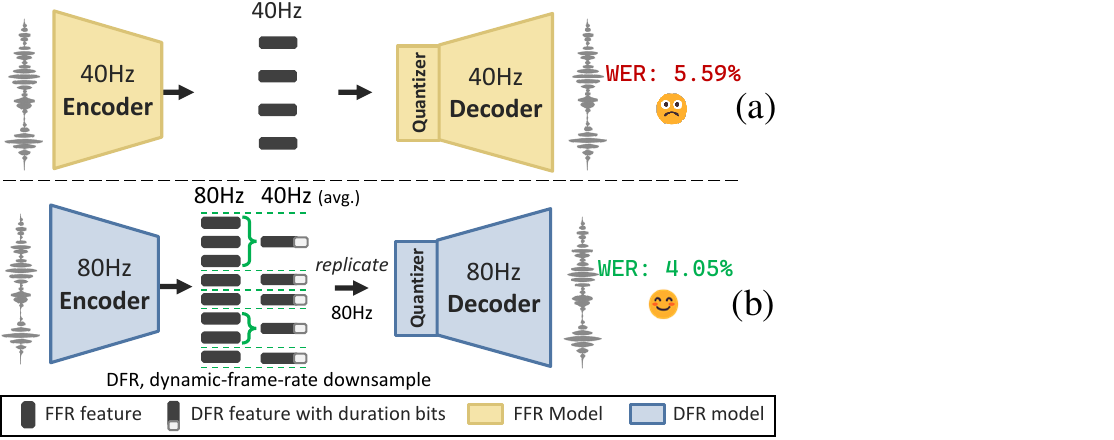}
  \vspace{-15pt}
    \caption{Comparison of:
  (a) conventional \SI{40}{Hz} fixed-rate model, 
    (b) {\ourmethod}-integrated model, which combines Melt-and-Cool training with ScheDFR for inference, achieving the lowest WER.}
  \label{fig:opening}
  \vspace{-15pt}
\end{figure}

To address these issues, we introduce {\ourmethod}, a novel plugin-style approach that breaks the limits of traditional FFR codecs by modifying existing FFR codec backbone models for both training and inference. 
{\ourmethod} accomplishes this by introducing two core techniques: \textit{ScheDFR} and \textit{Melt-and-Cool}.
As shown in Figure~\ref{fig:opening}, ScheDFR adaptively merges adjacent similar frames during inference, while the Melt-and-Cool adapts FFR models to DFR through post-training and fine-tuning.
Key advantages of {\ourmethod} are summarized as follows:
(1) \textbf{Enhanced performance under low frame rates and bitrates.}
To reduce frame rate, the {\ourmethod}-integrated model first encodes speech features at a higher frame rate, then adaptively selects downsampling schemes through optimization of a feature-space distortion metric, targeting better reconstruction quality preservation. This process inherently decouples duration and content information, enhancing information capacity over FFR codecs. 
(2)
\textbf{Frame rate controllability.}
{\ourmethod} enables \emph{a single model} to adapt to varying the target \emph{average} frame rate of an utterance at inference-time. 
Users can flexibly trade off between frame rate and performance based on downstream requirements or resource limitations.
(3)
\textbf{Fully unsupervised training.}
{\ourmethod} requires no external labels (e.g., transcripts, alignments, or speaker labels) at any training stage. This allows seamless scalability to diverse data domains.
(4)
\textbf{Orthogonal to codec backbones}.
Though {\ourmethod} is implemented on a single-codebook and non-disentangled general-purpose codec backbone~\cite{xin2024bigcodec} in this work, the core ScheDFR and Melt-and-Cool techniques are backbone-agnostic and compatible with various speech tokenizer architectures.

Our results demonstrate that when integrating {\ourmethod} with an \SI{80}{Hz} FFR backbone and targeting \SI{40}{Hz} during inference, the WER decreases by 32\% relative to a \SI{40}{Hz} FFR model with equivalent content bitrate, while other metrics also outperforms. Even compared to an FFR model with an identical total bitrate, the WER remains up to 8\% lower, with other metrics comparable. Furthermore, {\ourmethod} exhibits strong generalization: the same {\ourmethod}-integrated model achieves better reconstruction quality when targeting other frame rates like \SI{50}{Hz} or \SI{67}{Hz} compared to FFR models trained at those rates. In summary, our work presents a systematic exploration of temporal redundancy compression in speech codecs, proposing an effective solution that opens new possibilities for future speech tokenization and downstream applications.

\section{Method: CodecSlime}

We propose {\ourmethod}, a novel plugin-style method for neural codecs that enhances traditional FFR codecs by enabling DFR. 
It achieves flexible encoding and high-quality reconstruction at low frame rates through two key techniques:
(1) Schedulable Dynamic Frame Rate (ScheDFR, \cref{sec:method-schedfr}) that aggregates temporally similar features in the inference stage to enable low-loss compression;
(2) Melt-and-Cool training recipe (\cref{sec:method-meltcool}), a two-stage process that adapts the FFR backbone model to better handle DFRs. The backbone model for integrating {\ourmethod} will be elaborated in \cref{sec:method-struct}.

\subsection{Preliminary: backbone model}
\label{sec:method-struct}
\paragraph{Architecture} The backbone model employs a VQ-GAN~\cite{esser2021taming} architecture similar to BigCodec~\cite{xin2024bigcodec}: the generator consists of three components: an encoder, a quantizer, and a decoder, and a discriminator. Specifically, the encoder \( f_\mathrm{E}(\cdot) \) consists of convolutional neural networks (CNNs) and long short-term memory networks (LSTMs), and the decoder \( f_\mathrm{D}(\cdot) \) adopts the mirrored CNN structure of the encoder. The discriminator is made up of MPD and MS-STFT discriminators~\cite{encodec, kong2020hifigan}.
To further validate the backbone-agnostic property of our method, we evaluate it with two types of quantizers:  
(1) the vector quantizer (VQ)~\cite{VQVAE} used in BigCodec~\cite{xin2024bigcodec}, which maps a continuous feature $h \in \mathbb{R}^{d_\mathrm{h}}$ to the nearest entry in a trainable codebook. (2) the finite scalar quantizer (FSQ)~\cite{mentzer2024finite} used in systems like S3Tokenizer~\cite{cosyvoice2}, which quantizes a vector element-wise: it applies a per-channel bounding function to each dimension, then rounds the result to the nearest integer.

\paragraph{Training objective}
The training loss incorporates reconstruction loss and GAN loss, similar to BigCodec~\cite{xin2024bigcodec}. Both of the losses take part in all stages of \ourmethod's training. Specifically, 
(1) For the reconstruction loss, we employ the multi-scale mel-spectrogram loss~\cite{kumar2024high}, which computes L1 distances across multiple resolution scales.
(2) For the GAN loss, least-squares GAN objective and L1 feature matching loss~\cite{mao2017least} are adopted. 

\subsection{Schedulable Dynamic Frame Rate (ScheDFR)}
\label{sec:method-schedfr}

\subsubsection{Motivation} 
Speech is a sequence of phonemes (and silences), each with its duration and paralinguistic features, leading to non-uniform information density over time. For example, long vowels and consonants carry similar semantic information, yet vowels occupy longer durations. Existing FFR codecs, however, allocate more frames to phonemes with less temporal variation, such as silences or sustained vowels. To address this, we propose Schedulable Dynamic Frame Rate (ScheDFR), which compresses FFR speech feature sequences to a lower target frame rate while preserving reconstruction quality.

\subsubsection{Problem formalization}

ScheDFR inserts a schedulable downsampling module between the encoder and the quantizer. It takes as input the encoder outputs $\mathbf{h}^{T \times d_\mathrm{h}} = f_\mathrm{E}(\mathbf{x})$ and a target downsampling ratio $R_\mathrm{S}$, and outputs a segmentation scheme $\mathbf{s}^\ast = \{s_1, \ldots, s_{T'}\}$ with $T' = \lceil T / R_\mathrm{S} \rceil$. The procedure consists of downsampling and scheduling.

\paragraph{Downsampling}
Given a segment length sequence $\mathbf{s}=(s_1,\ldots,s_{T'})$ with $\sum_{i=1}^{T'} s_i=T$ and $1 \le s_i \le U$, define start indices $\sigma_1=1$ and $\sigma_{i+1}=\sigma_i+s_i$. The frame-averaged downsampling function
$f_\mathrm{down}: \mathbb{R}^{T \times d_\mathrm{h}} \times \mathbb{N}^{T'} \to \mathbb{R}^{T \times d_\mathrm{h}}$
is defined as
\vspace{-8pt}
\begin{equation}
    \forall i \in [1,T'],\;\forall t \in [\sigma_i, \sigma_i+s_i-1],\quad
    h'_t = \frac{1}{s_i}\sum_{j=\sigma_i}^{\sigma_i+s_i-1} h_j .
\vspace{-8pt}
\end{equation}
Thus the output $\mathbf{h}' = f_\mathrm{down}(\mathbf{h},\mathbf{s})$ has the same temporal length $T$, equivalent to compressing to $T'$ frames then upsampling. To preserve duration, each merged frame additionally stores $\lceil log_2 U \rceil$ bits, decoupling content and duration.

\paragraph{Scheduling}
Let $\mathcal{S}=\{\mathbf{s}\mid \sum_{i=1}^{T'} s_i=T,\,1\leq s_i\leq U\}$. The optimal segmentation is
\vspace{-6pt}
\begin{equation}
\begin{aligned}
    \mathbf{s}^\ast = \underset{\mathbf{s}\in\mathcal{S}}{\arg\max}\, \mathcal{J}(\hat{\mathbf{x}}', \mathbf{x}) \quad
        \textrm{where }\hat{\mathbf{x}}'=f_\mathrm{D}\left(\mathbf{h}'\right) 
\end{aligned}
\vspace{-6pt}
\label{eq:scheduling-theo}
\end{equation}
where $\mathcal{J}$ evaluates reconstruction quality. Since direct optimization of most reconstruction quality metrics is non-differentiable or intractable, we design a surrogate objective.

\paragraph{Surrogate objective}
The surrogate objective treat the quality of a segmentation $\mathbf{s}$ as the negative L2 distance between the original and the downsampled features: $\mathcal{J}_\mathrm{h}(\mathbf{h},\mathbf{s}) \;=\;  -\sum_{t=1}^{T} \lVert h_t - h'_t \rVert_2.$ 
The optimal segmentation is then obtained as $\mathbf{s}^\ast = \underset{\mathbf{s}\in\mathcal{S}}{\arg\max}\;\mathcal{J}_\mathrm{h}(\mathbf{h},\mathbf{s}).$

\vspace{-6pt}
\subsubsection{DP-based downsample scheduler}
\label{sec:dp-ds-sche}
We propose to solve the optimization problem of the surrogate objective $\mathcal{J}_{\mathrm h}$ 
with a dynamic-programming (DP) algorithm. 
Let $d[j,i]$ denote the maximum objective value when the first $j$ frames have been downsampled into $i$ frames (i.e., split into $i$ segments), and $L(j,s) = \sum_{t=j-s+1}^{j} \left\| h_t - \frac{1}{s}\sum_{k=j-s+1}^{j} h_k \right\|_2 .$
The DP maximizes $\mathcal{J}_{\mathrm h}$ with the DP equation below:
\vspace{-6pt}
\begin{equation}
d[0,0]=0;\qquad
d[j,i]=
\max_{1\le s\le U}
        \Bigl\{\,d[j-s,i-1] - L(j, s)\Bigr\},
\vspace{-6pt}
\end{equation}
where $L$ values can be pre-processed.
The optimum is \(d[T,T']\).  
Thus, through the downsample scheduler, ScheDFR adaptively identifies and compresses temporal redundancy in speech signals, achieving high-quality compression and reconstruction at low frame rates.

\subsection{Melt-and-Cool training recipe}
\label{sec:method-meltcool}

\begin{figure}[tb]
  \centering
    \includegraphics[width=0.49\textwidth, trim=0cm 5.4cm 10cm 0cm, clip]{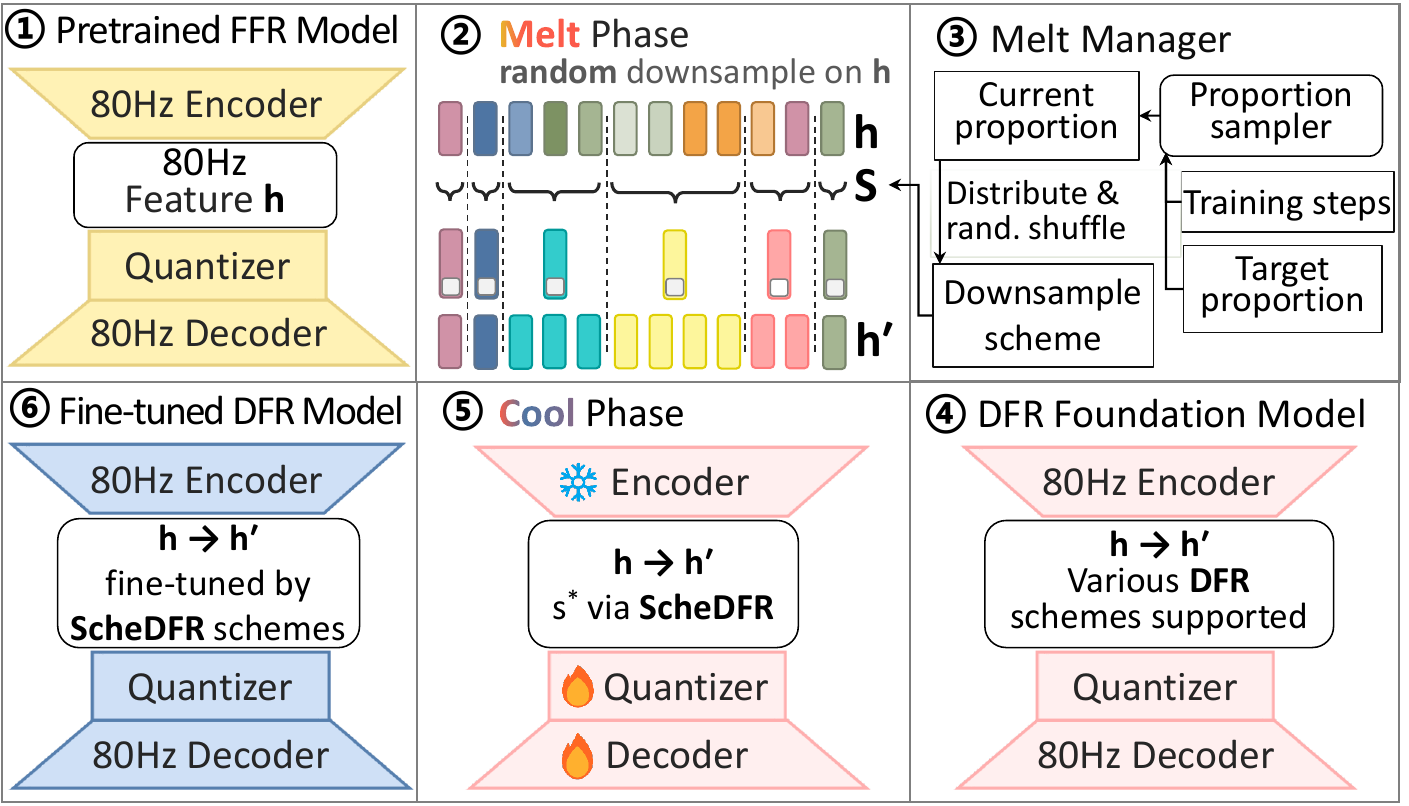}
  \vspace{-17pt}
    \caption{Overview of the Melt-and-Cool training recipe.}
  \label{fig:melt-and-cool}
  \vspace{-12pt}
\end{figure}

\paragraph{Motivation} Although ScheDFR clearly outperforms fixed-rate downsampling, applying it directly to an FFR backbone still leads to high WER. The main reason is that the backbone has never been trained to handle merged-frame encoding and decoding. We therefore propose a two-stage training recipe, Melt-and-Cool (Fig.~\ref{fig:melt-and-cool}). Melt post-trains the model with random downsampling to cover diverse downsampled inputs, and Cool fine-tunes it with optimal ScheDFR schemes for specific $R_\mathrm{S}$ and $U$. Following prior work, both stages use fixed-length randomly cropped speech segments.

\paragraph{Post-training (Melt)} We start from a pretrained FFR model (\circled{1}) and randomly downsample the encoded features (\circled{2}). A Melt manager (\circled{3}) increases the downsampling strength over time by controlling the proportion of segments of different lengths while keeping the diversity of downsampling schemes. This yields a DFR foundation model (\circled{4}) that is robust to many downsampling schemes.

\paragraph{Fine-tuning (Cool)} 
In the Cool stage, we fine-tune the foundation model under ScheDFR for the target $R_\mathrm{S}$ and $U$ (\circled{5}). The DP-based scheduler is run inside the forward process to obtain the optimal scheme for each training segment on the fly. The encoder is frozen, and only the quantizer and decoder are updated. The resulting fine-tuned DFR model (\circled{6}) is thus tailored to ScheDFR while retaining the robustness gained from Melt.

\section{Experimental setup}
\subsection{Experimental setup}
\paragraph{Backbone model} 
We train the backbone model using the full LibriSpeech~\cite{panayotov2015librispeech} training set (960 hours of \SI{16}{kHz} 16-bit audio). Model parameters include: token frequency of \SI{80}{Hz}; feature dimension $d_\mathrm{h}=1024$ for $\mathbf{h}$; VQ codebook size of 8192 and FSQ codebook size of 18225 with projected dimension 8. All models are trained on 2 NVIDIA A800 GPUs. We use AdamW~\cite{loshchilov2017adamw} optimizer with moving average coefficients $\beta_1=0.8$ and $\beta_2=0.9$, and a linearly decaying learning rate from $10^{-4}$ to $10^{-5}$ with 1000 warm-up steps. The backbone model is trained for 1.2M steps. The training code is adopted from the official implementation of BigCodec\footnote{\scriptsize{https://github.com/Aria-K-Alethia/BigCodec}}~\cite{xin2024bigcodec}.

\paragraph{Integrating \ourmethod}
The main experiment of this paper is to downsample the \SI{80}{Hz} backbone codec to \SI{40}{Hz}. 
The Melt stage still uses the same learning rate configuration as the backbone.
In the Cool stage, downsampling schemes are computed using the DP-based downsample scheduler with $R_\mathrm{S}=2, U=4$. The learning rate decays from $4\times 10^{-5}$ to $10^{-5}$. Other settings remain the same as in previous stages. Each stages is trained for $\approx$100k steps.

\paragraph{Baselines}
To validate our method, we compare \ourmethod-integrated models with \SI{40}{Hz} fixed-frame-rate models with similar content bitrate and total bitrate (i.e. content plus duration). These models are trained using the official implementation of BigCodec~\cite{xin2024bigcodec}. \textbf{BigCodec-VQ\textsubscript{8k}} uses the original quantizer and codebook size of BigCodec. \textbf{BigCodec-VQ\textsubscript{18k}} uses a codebook size of 18225. \textbf{BigCodec-FSQ\textsubscript{18k}} uses FSQ with 18225 codes, and \textbf{BigCodec-FSQ\textsubscript{84k}} enlarges the codebook to 84375.

Additionally, we compare our method with multiple codecs using their official checkpoints:
\textbf{EnCodec}~\cite{encodec}, \textbf{LLM-Codec}~\cite{yang2024uniaudio15}, \textbf{SNAC}~\cite{Siuzdak_SNAC_Multi-Scale_Neural_2024},
\textbf{TFC}~\cite{zhang2025unlocking} and \textbf{VARSTok}~\cite{zheng2025say}.
For EnCodec, we use the model operating at \SI{1.5}{kbps}. 
LLM-Codec employs its official checkpoint. 
For SNAC, the \SI{0.98}{kbps} model trained on pure speech is used.
For TFC, we use the fine-level frame rate (\SI{75}{Hz}) model, and only utilize the first codebook for fair comparison.
For VARSTok, the $\tau=0.8$ version is employed.
When evaluating models trained on \SI{24}{khz} speech, we upsample input audio to \SI{24}{kHz} before encoding and subsequently downsample to \SI{16}{kHz}.

\paragraph{Testset} We use \textbf{UniCATS testset B}~\cite{du2024unicats}, a subset of LibriTTS~\cite{libritts} test-clean, which contains 500 utterances from 37 unseen speakers.

\paragraph{Metrics} 
Token properties include \textbf{frame rate}, \textbf{codebook size}, and \textbf{bitrate}, decoupled into \textit{content} and \textit{duration} components. Reconstruction quality is assessed through objective measures such as \textbf{WER} (using NeMo ASR\footnote{\scriptsize{https://huggingface.co/nvidia/stt\_en\_fastconformer\_transducer\_large}}),
\textbf{STOI}\footnote{\scriptsize{https://github.com/mpariente/pystoi}}, \textbf{PESQ}\footnote{\scriptsize{https://github.com/ludlows/PESQ}}, \textbf{ViSQOL}~\cite{visqolv3}, \textbf{UTMOS}~\cite{saeki22c_interspeech}, and speaker similarity measured by \textbf{SECS} (Resemblyzer\footnote{\scriptsize{https://github.com/resemble-ai/Resemblyzer}}). Additionally, we also conduct a subjective \textbf{MUSHRA}~\cite{series2014method} 
test (ITU-R BS.1534) with 21 participants, with each evaluates 7 utterances (from the testset) across 9 systems: one ground truth reference, one anchor (SNAC), and 7 models (VARSTok, BigCodec-VQ\textsubscript{8k}, BigCodec-VQ\textsubscript{18k}, CodecSlime-VQ\textsubscript{8k}, BigCodec-FSQ\textsubscript{18k}, BigCodec-FSQ\textsubscript{84k}, CodecSlime-FSQ\textsubscript{18k}). The candidate number is limited to 9 to avoid excessive cognitive load and ensure reliable listener ratings within a practical evaluation time. Scores are collected on a 0–100 scale in blind listening.

\begin{table*}[t]
\caption{
Reconstruction results of \ourmethod and all baselines listed. Frame Rate (Hz): Single-layer (f×1), multi-layer with identical frequencies (f×n), or different frequencies (f1+f2+...). Rows shown in gray font are codebook-enlarged FFR baselines.}
\vspace*{-8pt}
\centering
\label{tab:main}
\renewcommand{\arraystretch}{1.2}
\setlength{\tabcolsep}{4pt}
\scalebox{0.9}{
\begin{tabular}{lcc|cc|ccccccc}
\toprule
\multirow{2}{*}{\textbf{Model}} & \multirow{2}{*}{\makecell{\textbf{Frame} \\ \textbf{Rate} (Hz)}} & \multirow{2}{*}{\makecell{\textbf{Codebook} \\ \textbf{Size}}} & \multicolumn{2}{c|}{\textbf{Bitrate} (kbps)} & \multirow{2}{*}{\makecell{\textbf{WER} \\(\%)$\downarrow$}} & \multirow{2}{*}
{\textbf{STOI$\uparrow$}} & \multirow{2}{*}{\textbf{PESQ$\uparrow$}} & \multirow{2}{*}
{\textbf{SECS$\uparrow$}} & \multirow{2}{*}{\textbf{ViSQOL$\uparrow$}} & \multirow{2}{*}{\textbf{UTMOS$\uparrow$}} & \multirow{2}{*}{\textbf{MUSHRA$\uparrow$}}\\
\cline{4-5}
 & & & \textbf{Content} & \textbf{Duration} & & & & & & \\
\midrule
Ground Truth
& -  & -      & 256    & -    & 1.16 & 1.000 & 4.50 & 1.000 & 5.00 & 4.14 & {\color{black}91.60{\scriptsize $\pm$1.87}}\\
\hline
EnCodec~\cite{encodec}
& $75 \times 2$ & 1024   & 1.50 & -  & 4.13  & 0.856 & 1.59 & 0.894 & 3.74 & 1.64 & - \\
LLM-Codec
~\cite{yang2024uniaudio15}
& {\footnotesize 8+17+33} & {$\approx$32000}  & 0.85 & -    & 6.25  & 0.879 & 1.82 & 0.917 & 3.77 & 2.71 & {\color{black}-}\\
SNAC~\cite{Siuzdak_SNAC_Multi-Scale_Neural_2024} & {\footnotesize 12+23+47} & 4096 & 0.98 & - & 2.67 & 0.897 & 1.92 & 0.905 & 3.76 & 3.16 & {\color{black}52.70{\scriptsize$\pm$2.61}} \\
{\color{black}TFC-fine~\cite{zhang2025unlocking}} & {\color{black}$75 \times 1$} & {\color{black}1024} & {\color{black}0.75} & {\color{black}-} & {\color{black}14.56} & {\color{black}0.812} & {\color{black}1.36} & {\color{black}0.826} & {\color{black}3.37} & {\color{black}2.04} & {\color{black}-} \\
{\color{black}VARSTok($\tau=0.8$)}~\cite{zheng2025say} & {\color{black}$37.8 \times 1$} & {\color{black}4096} & {\color{black}0.53} & {\color{black}-} & {\color{black}7.26} & {\color{black}0.879} & {\color{black}1.81} & {\color{black}0.883} & {\color{black}4.01} & {\color{black}3.79} & {\color{black}77.37{\scriptsize$\pm$2.24}} \\
\hline
BigCodec-VQ$_\textrm{8k}$~\cite{xin2024bigcodec} & $40 \times 1$ & 8192 & 0.52 & - & 4.89 & 0.896 & 1.98 & 0.932 & 3.91 & 4.00 & {\color{black}73.45{\scriptsize$\pm$2.81}} \\
\color{gray}BigCodec-VQ$_\textrm{18k}$
& \color{gray}$40 \times 1$ & \color{gray}18225  & \color{gray}0.57 & \color{gray}-    & \color{gray}4.64  & \color{gray}0.899 & \color{gray}\textbf{2.02} & \color{gray}0.932 & \color{gray}3.93 & \color{gray}\textbf{4.02} & \color{gray}79.68{\scriptsize $\pm$1.63}\\
\hdashline
\ourmethod-VQ$_\textrm{8k}$ & $40 \times 1$ & 8192 & 0.52 & 0.08 & \textbf{4.25} & \textbf{0.903} & \textbf{2.02} & \textbf{0.935} & \textbf{3.96} &  \textbf{4.02} & {\color{black}\textbf{84.01}{\scriptsize $\pm$1.59}}\\
\hline
BigCodec-FSQ$_\textrm{18k}$
& $40 \times 1$ & 18225  & 0.57 & -    & 5.59  & 0.891 & 1.90 & 0.913 & 3.85 & 3.93 & {\color{black}74.42{\scriptsize $\pm$2.14}} \\
\color{gray} BigCodec-FSQ$_\textrm{84k}$
& \color{gray}$40 \times 1$ & \color{gray}84375  & \color{gray}0.65 & \color{gray}-    & \color{gray}4.12  & \color{gray}0.901 & \color{gray}\textbf{2.03} & \color{gray}\textbf{0.918} & \color{gray}\textbf{3.91} & \color{gray}3.99 & \color{gray}77.03{\scriptsize $\pm$1.92} \\
\hdashline
{\color{black}CodecSlime-FSQ$_\textrm{18k}$} & {\color{black}$40 \times 1$} & {\color{black}18225} & {\color{black}0.57} & {\color{black}0.08} & {\color{black}\textbf{3.80}} & {\color{black}\textbf{0.904}} & {\color{black}2.00} & {\color{black}\textbf{0.918}} & {\color{black}\textbf{3.91}} & {\color{black}\textbf{4.05}} & {\color{black}\textbf{81.24}{\scriptsize$\pm$1.88}} \\
\bottomrule
\end{tabular}
}
\vspace*{-12pt}
\end{table*}

\subsection{Main results}

Table~\ref{tab:main} compares all baselines, where our {\ourmethod}, implemented as a $2\times$ dynamic downsampling plugin atop \SI{80}{Hz} backbones with VQ\textsubscript{8\text{k}} and FSQ\textsubscript{18\text{k}} respectively, consistently achieves the best results. 
In the VQ setting, {\ourmethod} reduces WER to 4.25\%, outperforming both BigCodec-VQ\textsubscript{8k} and BigCodec-VQ\textsubscript{18k} while also yielding higher STOI and ViSQOL, with subjective MUSHRA scores showing a significant preference for {\ourmethod}. 
In the FSQ setting, {\ourmethod} achieves the lowest WER of 3.80\%, showing a substantial gain over its FFR baseline BigCodec-FSQ\textsubscript{18k} (5.59\%), while maintaining comparable or superior scores across other metrics such as STOI, UTMOS, and MUSHRA. 
Even against the stronger high-capacity baseline BigCodec-FSQ\textsubscript{84k}, whose content bitrate matches {\ourmethod} in total bitrate, our method achieves a lower WER with comparable perceptual metrics, demonstrating robust semantic and acoustic preservation. 
These results confirm the effectiveness of {\ourmethod} in compressing temporal redundancy while preserving intelligibility and perceptual quality regardless of the backbone quantizer.

\subsection{Evaluation of generalization ability}
\begin{figure}[tp]
\centering
    \includegraphics[width=0.95\linewidth]{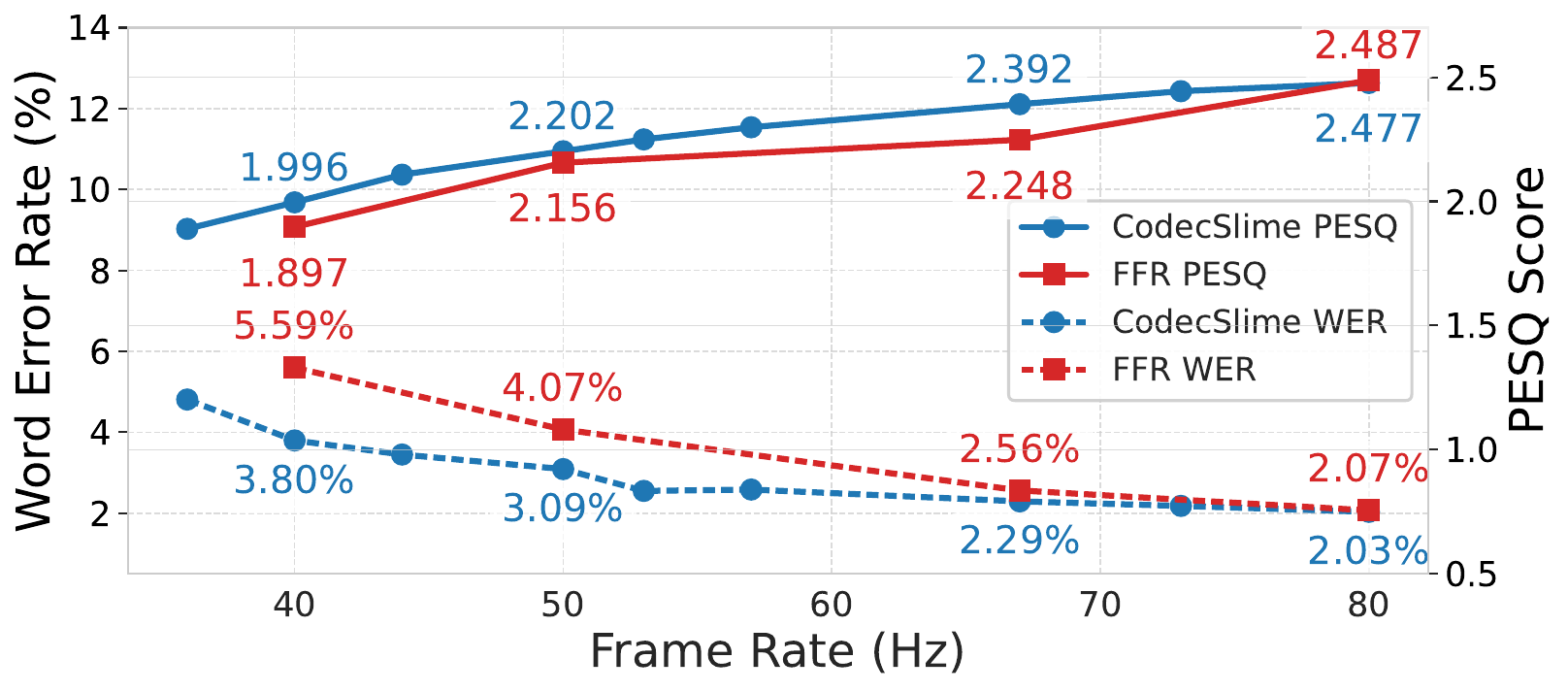}
    \vspace*{-10pt}
    \captionof{figure}{WER and PESQ across frame rates for {\ourmethod} (one single model) and FFR baselines (different models for different frame rates). Although all models shown in the figure use FSQ, VQ-version models also show similar trends.}
    \label{fig:wer_stoi_vs_fr}
\vspace*{-15pt}
\end{figure}

Figure~\ref{fig:wer_stoi_vs_fr} illustrates how WER and PESQ vary across different target inference frame rates for {\ourmethod} and the fixed-rate FFR baseline. In this experiment, {\ourmethod} uses \emph{the same model} fine-tuned under \SI{40}{Hz} ScheDFR, while the FFR baselines are separately trained models at each specific frame rate (\SI{40}{Hz}, \SI{50}{Hz}, \SI{67}{Hz} and \SI{80}{Hz}). All models use FSQ with 18225 codes.
The figure reveals a consistent trend: as the inference frame rate increases, WER decreases and PESQ increases for both {\ourmethod} and FFR models. {\ourmethod} consistently outperforms FFR across the range from \SI{40}{Hz} to \SI{80}{Hz}. 
These results demonstrate that {\ourmethod} generalizes effectively across a wide range of frame rates, despite being trained only once at a single dynamic average frame rate. In contrast, FFR requires retraining separate models for each target frame rate, yet still fails to match {\ourmethod}’s performance. This confirms {\ourmethod}'s strong generalization and flexibility on inference frame rate.

We also evaluated {\ourmethod} on unseen languages using a subset of MLS~\cite{MLS} testset. It also shows better reconstruction quality compared to FFR baselines (improving WER by over 17\% while keeping other metrics comparable), showing strong generalization ability on out-of-domain inputs. 

\subsection{Ablation studies}
\label{sec:res-ablation}
\begin{table}[tp]
\begin{minipage}{0.46\linewidth}
\caption{Ablation of ScheDFR}
\centering
\renewcommand{\arraystretch}{1.2}
\setlength{\tabcolsep}{2pt}
\vspace*{-8pt}
\scalebox{0.6}{
\begin{tabular}{lcHcHHc}
\toprule
\textbf{Model} (FSQ) & 
\textbf{WER}(\%)$\downarrow$ & \textbf{STOI$\uparrow$} & \textbf{PESQ$\uparrow$} & \textbf{SECS$\uparrow$} & \textbf{MCD$\downarrow$} & \textbf{UTMOS$\uparrow$} \\
\midrule
DFR Fdn. Model      & 5.02 & 0.892 & 1.85 & 0.900 & 1.81 & 3.92 \\
\quad + ScheDFR           
& \textbf{4.42} & \textbf{0.900} & \textbf{1.91} & \textbf{0.913} & \textbf{1.78} & \textbf{3.97} \\
\midrule
{Fine-tuned Model}          & 4.65 & 0.893 &  1.82 & 0.909 & xxx & 3.87 \\
\quad + ScheDFR           
& \textbf{3.80} & \textbf{0.904} & \textbf{2.00} & \textbf{0.918} & \textbf{} & \textbf{4.05} \\
\bottomrule
\end{tabular}
}
\label{tab:ablation_schedfr}
\end{minipage}
\vspace*{-9pt}
\hfill
\begin{minipage}{0.52\linewidth}
\caption{Ablation of Melt-and-Cool}
\centering
\renewcommand{\arraystretch}{1.5}
\setlength{\tabcolsep}{2pt}
\vspace*{-8pt}
\scalebox{0.6}{
\begin{tabular}{lcHcHHc}
\toprule
\textbf{Model} (FSQ) & 
\textbf{WER}(\%)$\downarrow$ & \textbf{STOI$\uparrow$} & \textbf{PESQ$\uparrow$} & 
\textbf{SECS$\uparrow$} & \textbf{MCD$\downarrow$} & \textbf{UTMOS$\uparrow$} \\
\midrule
FFR Backbone             
& 18.59 & 0.761 & 1.18 & 0.860 & 2.81 & 2.57 \\
\quad + Cool (w/o Melt)  
& 8.22  & 0.855 & 1.37 & 0.911 & 2.11 & 3.31 \\
\quad + Melt-and-Cool    
& \textbf{3.80} & \textbf{0.904} & \textbf{2.00} & \textbf{0.918} & \textbf{} & \textbf{4.05} \\
\bottomrule
\end{tabular}
}
\label{tab:ablation_meltcool}
\end{minipage}
\vspace*{-5pt}
\end{table}

\paragraph{On ScheDFR} 
Table~\ref{tab:ablation_schedfr} reports results of compressing \SI{80}{Hz} features to \SI{40}{Hz} during inference. When using a fixed pattern (merging every two frames), both the DFR foundation model and the fine-tuned model yield suboptimal performance, with WER 5.02\% and 4.65\% and degraded perceptual scores. By contrast, equipping the same models with ScheDFR significantly improves all metrics, reducing WER to 4.42\% and 3.80\% respectively while also increasing PESQ and UTMOS. These results confirm that adaptive frame merging is essential for effective temporal redundancy compression.

\paragraph{On Melt-and-Cool} 
Table~\ref{tab:ablation_meltcool} evaluates the contribution of the two-stage training recipe, with ScheDFR applied in all cases. Starting from the FFR backbone, direct inference under DFR leads to poor performance (WER 18.59\%). Introducing only the Cool stage improves the model (WER 8.22\%, UTMOS 3.31), but still lags far behind the full Melt-and-Cool process. With both stages combined, WER is reduced to 3.80\% and perceptual scores also reach the best overall quality, demonstrating the necessity of the Melt-and-Cool recipe.

\section{Conclusion}
We presented {\ourmethod}, a plugin-style method that endows FFR neural speech codecs with DFR control.
It combines an inference-time technique, ScheDFR, and a post-training recipe, Melt-and-Cool, to compress temporal redundancy and decouple content and duration information. Both are fully unsupervised and architecture-agnostic.
Integrated with a typical VQ-GAN backbone and inferred at \SI{40}{Hz}, {\ourmethod} yields substantially lower WER (up to 32\% relatively) than FFR models of similar bitrate, while a single model generalizes across higher frame rates without retraining.
These findings provide systematic evidences that temporal redundancy in codecs can be compressed effectively through DFR, charting a path toward more efficient speech tokenization and downstream applications.

\section{Acknowledgement}
This work has been supported by the China NSFC Project (No. 92370206) and the Key Research and Development Program of Jiangsu Province, China (No.BE2022059). We also thank Hanglei Zhang and Borui Zhang for their contribution to this work.

\bibliographystyle{IEEEbib}
\bibliography{main}

@string{interspeech = "Proc. ISCA Interspeech"}

@string{aaai = "Proc. AAAI"}

@string{icassp = "Proc. IEEE ICASSP"}

@string{iclr = "Proc. ICLR"}

@string{ieee-acm-taslp = "IEEE/ACM Trans. ASLP."}

@string{iccv = "Proc. ICCV"}

@string{nips = "Proc. NeurIPS"}

@string{iccv = "Proc. IEEE/CVF ICCV"}

@inproceedings{du2024unicats,
  title={{UniCATS: A Unified Context-Aware Text-to-Speech Framework with Contextual VQ-Diffusion and Vocoding}},
  author = {Du, Chenpeng and Guo, Yiwei and Shen, Feiyu and others},
  booktitle=aaai,
  volume={38},
  pages={17924--17932},
  year={2024}
}

@article{kong2020hifigan,
  title={{Hifi-GAN: Generative Adversarial Networks for Efficient and High Fidelity Speech Synthesis}},
  author = {Kong, Jungil and Kim, Jaehyeon and Bae, Jaekyoung},
  journal=nips,
  year={2020}
}

@article{hsu2021hubert,
  title={{HuBERT: Self-Supervised Speech Representation Learning by Masked Prediction of Hidden Units}},
  author = {Hsu, Wei-Ning and Bolte, Benjamin and Tsai, Yao-Hung and others},
  journal=ieee-acm-taslp,
  volume={29},
  pages={3451--3460},
  year={2021},
  publisher={IEEE}
}

@inproceedings{libritts,
  author = {Heiga Zen and Viet Dang and Rob Clark and others},
  title={{LibriTTS: A Corpus Derived from LibriSpeech for Text-to-Speech}},
  year=2019,
  booktitle=interspeech,
  pages={1526--1530},
  doi={10.21437/Interspeech.2019-2441},
  issn={2958-1796}
}

@article{baevski2020wav2vec,
  title={{wav2vec 2.0: A Framework for Self-Supervised Learning of Speech Representations}},
  author = {Baevski, Alexei and Zhou, Yuhao and Mohamed, Abdelrahman and others},
  journal=nips,
  year={2020}
}

@inproceedings{MLS,
  author = {Vineel Pratap and Qiantong Xu and Anuroop Sriram and others},
  title={{MLS: A Large-Scale Multilingual Dataset for Speech Research}},
  year=2020,
  booktitle=interspeech,
  pages={2757--2761},
  doi={10.21437/Interspeech.2020-2826},
  issn={2958-1796}
}

@article{
encodec,
title={{High Fidelity Neural Audio Compression}},
author={Alexandre D{\'e}fossez and Jade Copet and Gabriel Synnaeve and others},
journal={TMLR},
issn={2835-8856},
year={2023},
}

@article{kumar2024high,
  title={{High-Fidelity Audio Compression with Improved RVQGAN}},
  author = {Kumar, Rithesh and Seetharaman, Prem and Luebs, Alejandro and others},
  journal=nips,
  volume={36},
  year={2024}
}

@ARTICLE{valle,
  author = {Chen, Sanyuan and Wang, Chengyi and Wu, Yu and others},
  journal=ieee-acm-taslp, 
  title={{Neural Codec Language Models are Zero-Shot Text to Speech Synthesizers}}, 
  year={2025},
  volume={},
  number={},
  pages={1-15},
  doi={10.1109/TASLPRO.2025.3530270}
}

@article{kyutai2024moshi,
  title={{Moshi: A Speech-Text Foundation Model for Real-Time Dialogue}},
  author={D{\'e}fossez, Alexandre and Mazar{\'e}, Laurent and Orsini, Manu and others},
  journal={arXiv preprint arXiv:2410.00037},
  year={2024}
}

@inproceedings{ticodec,
  title={{Fewer-Token Neural Speech Codec with Time-Invariant Codes}},
  author = {Ren, Yong and Wang, Tao and Yi, Jiangyan and others},
  booktitle=icassp,
  year={2024},
}

@inproceedings{
zhang2024speechtokenizer,
title={{SpeechTokenizer: Unified Speech Tokenizer for Speech Language Models}},
author = {Xin Zhang and Dong Zhang and Shimin Li and others},
booktitle=iclr,
year={2024},
}

@article{zeghidour2021soundstream,
  title={{SoundStream: An End-to-End Neural Audio Codec}},
  author = {Zeghidour, Neil and Luebs, Alejandro and Omran, Ahmed and others},
  journal=ieee-acm-taslp,
  volume={30},
  pages={495--507},
  year={2021},
  publisher={IEEE}
}

@article{xin2024bigcodec,
    title={{BigCodec}: {Pushing the Limits of Low-Bitrate Neural Speech Codec}},
  author = {Xin, Detai and Tan, Xu and Takamichi, Shinnosuke and others},
  journal={arXiv preprint arXiv:2409.05377},
  year={2024}
}

@article{borsos2023audiolm,
  title={{AudioLM}: {A Language Modeling Approach to Audio Generation}},
  author={Borsos, Zal{\'a}n and Marinier, Rapha{\"e}l and Vincent, Damien and others},
  journal=ieee-acm-taslp,
  volume={31},
  pages={2523--2533},
  year={2023},
  publisher={IEEE}
}

@article{VQVAE,
  title={{Neural Discrete Representation Learning}},
  author = {Van Den Oord, Aaron and Vinyals, Oriol and others},
  journal=nips,
  volume={30},
  year={2017}
}

@inproceedings{
mentzer2024finite,
title={{Finite Scalar Quantization: VQ-VAE Made Simple}},
author = {Fabian Mentzer and David Minnen and others},
booktitle=iclr,
year={2024},
}

@inproceedings{
yang2024uniaudio15,
title={{UniAudio} 1.5: {Large Language Model-Driven Audio Codec is A Few-Shot Audio Task Learner}},
author = {Dongchao Yang and Haohan Guo and Yuanyuan Wang and others},
booktitle=nips,
year={2024},
}

@article{brown2020language,
  title={Language models are few-shot learners},
  author={Brown, Tom and Mann, Benjamin and Ryder, Nick and others},
  journal={Advances in neural information processing systems},
  volume={33},
  pages={1877--1901},
  year={2020}
}

@article{Siuzdak_SNAC_Multi-Scale_Neural_2024,
      title={{SNAC}: {Multi-Scale Neural Audio Codec}}, 
      author = {Hubert Siuzdak and Florian Grötschla and Luca A. Lanzendörfer},
      year={2024},
        journal={arXiv preprint arXiv:2410.14411},
}

@inproceedings{esser2021taming,
  title={{Taming Transformers for High-Resolution Image Synthesis}},
  author = {Esser, Patrick and Rombach, Robin and Ommer, Bjorn},
  booktitle=iccv,
  pages={12873--12883},
  year={2021}
}

@article{wu2024ts3codec,
      title={{TS3-Codec}: {Transformer-Based Simple Streaming Single Codec}}, 
      author = {Haibin Wu and Naoyuki Kanda and Sefik Emre Eskimez and others},
      year={2024},
    journal={arXiv preprint arXiv:2411.18803}
}

@article{dieleman2021variable,
  title={{Variable-Rate Discrete Representation Learning}},
  author = {Dieleman, Sander and Nash, Charlie and Engel, Jesse and others},
  journal={arXiv preprint arXiv:2103.06089},
  year={2021}
}

@inproceedings{
baade2024syllablelm,
title={Syllable{LM: Learning Coarse Semantic Units for Speech Language Models}},
author={Alan Baade and Puyuan Peng and David Harwath},
booktitle=iclr,
year={2025},
}

@inproceedings{
cho2024sylber,
title={{Sylber: Syllabic Embedding Representation of Speech from Raw Audio}},
author={Cheol Jun Cho and Nicholas Lee and Akshat Gupta and others},
booktitle=iclr,
year={2025},
}

@article{cuervo2022variable,
  title={{Variable-Rate Hierarchical CPC Leads to Acoustic Unit Discovery in Speech}},
  author={Cuervo, Santiago and Lancucki, Adrian and Marxer, Ricard and others},
  journal=nips,
  year={2022}
}

@article{cosyvoice2,
      title={{CosyVoice} 2: {Scalable Streaming Speech Synthesis with Large Language Models}}, 
      author = {Zhihao Du and Yuxuan Wang and Qian Chen and others},
      year={2024},
journal={arXiv preprint arXiv:2412.10117}
}

@article{wang2025spark,
  title={Spark-tts: An efficient llm-based text-to-speech model with single-stream decoupled speech tokens},
  author={Wang, Xinsheng and Jiang, Mingqi and others},
  journal={arXiv preprint arXiv:2503.01710},
  year={2025}
}

@article{denes1963statistics,
  title={On the statistics of spoken English},
  author={Denes, Peter B},
  journal={The Journal of the Acoustical Society of America},
  year={1963},
  publisher={Acoustical Society of America}
}

@inproceedings{van2017information,
  title={On the information rate of speech communication},
  author={Van Kuyk, Steven and Kleijn, W Bastiaan and Hendriks, Richard C},
  booktitle=icassp,
  pages={5625--5629},
  year={2017},
  organization={IEEE}
}

@inproceedings{panayotov2015librispeech,
  title={Librispeech: an asr corpus based on public domain audio books},
  author={Panayotov, Vassil and Chen, Guoguo and Povey, Daniel and others},
  booktitle=icassp,
  pages={5206--5210},
  year={2015},
  organization={IEEE}
}

@article{loshchilov2017adamw,
  title={Decoupled weight decay regularization},
  author={Loshchilov, Ilya and Hutter, Frank},
  journal={arXiv preprint arXiv:1711.05101},
  year={2017}
}

@article{guo2025recent,
  title={Recent Advances in Discrete Speech Tokens: A Review},
  author={Guo, Yiwei and Li, Zhihan and Wang, Hankun and others},
  journal={arXiv preprint arXiv:2502.06490},
  year={2025}
}

@inproceedings{mao2017least,
  title={Least squares generative adversarial networks},
  author={Mao, Xudong and Li, Qing and Xie, Haoran and Lau, Raymond YK and Wang, Zhen and Paul Smolley, Stephen},
  booktitle={Proc. ICCV},
  year={2017}
}

@inproceedings{saeki22c_interspeech,
  title     = {{UTMOS: UTokyo-SaruLab System for VoiceMOS Challenge 2022}},
  author    = {Takaaki Saeki and Detai Xin and Wataru Nakata and others},
  year      = {2022},
  booktitle = {Interspeech 2022},
  pages     = {4521--4525},
  doi       = {10.21437/Interspeech.2022-439},
  issn      = {2958-1796},
}

@inproceedings{visqolv3,
  title={ViSQOL v3: An open source production ready objective speech and audio metric},
  author={Chinen, Michael and Lim, Felicia SC and Skoglund, Jan and others},
  booktitle={Proc. QoMEX},
  pages={1--6},
  year={2020},
  organization={IEEE}
}

@article{zhang2025unlocking,
  title={Unlocking Temporal Flexibility: Neural Speech Codec with Variable Frame Rate},
  author={Zhang, Hanglei and Guo, Yiwei and Li, Zhihan and others},
  journal={arXiv preprint arXiv:2505.16845},
  year={2025}
}

@article{series2014method,
  title={Method for the subjective assessment of intermediate quality level of audio systems},
  author={Series, B},
  journal={International Telecommunication Union Radiocommunication Assembly},
  volume={2},
  year={2014}
}

@article{zheng2025say,
  title={Say More with Less: Variable-Frame-Rate Speech Tokenization via Adaptive Clustering and Implicit Duration Coding},
  author={Zheng, Rui-Chen and Liu, Wenrui and Du, Hui-Peng and others},
  journal={arXiv preprint arXiv:2509.04685},
  year={2025}
}

\end{document}